# Electronic structure and magnetism of the Hund's insulator CrI$_3$

Tianye Yu, Rui Liu, Huican Mao, Xiaobo Ma, Guangwei Wang, Zhihong Yuan, Pengyu Zheng, Yiran Peng, and Zhiping Yin[*]

Department of Physics and Center for Advanced Quantum Studies, Beijing Normal University, Beijing 100875, China

**Abstract**

CrI$_3$ is a two-dimensional ferromagnetic van der Waals material with a charge gap of 1.1-1.2 eV. In this study, the electronic structure and magnetism of CrI$_3$ are investigated by using density functional theory and dynamical mean-field theory. Our calculations successfully reproduce a charge gap of ~1.1 eV in the paramagnetic state when a Hund's coupling $J_H = 0.7$ eV is included with an on-site Hubbard $U = 5$ eV. In contrast, with a large $U$ value of 8 eV and negligible Hund's coupling $J_H$, CrI$_3$ is predicted to be a moderately correlated metal in the paramagnetic state. We conclude that CrI$_3$ is a Mott-Hund's insulator due to the half-filled configuration of the Cr 3$d$ $t_{2g}$ orbitals. The Cr 3$d$ $e_g$ orbitals are occupied by approximately one electron, which leads to strong valence fluctuations so that the Cr 3$d$ orbitals cannot be described by a single state. Moreover, at finite temperature, the calculated ordered static magnetic moment in the ferromagnetic state is significantly larger in the $R\bar{3}$ phase than in the $C2/m$ phase. This observation indicates that the structural phase transition from the $C2/m$ phase to the $R\bar{3}$ phase with decreasing temperature is driven by ferromagnetic spin fluctuations.

## I. Introduction

Since the discovery of graphene[1-3], the field of two-dimensional (2D) materials has attracted widespread attention because of their outstanding mechanical, optical, and electrical properties, etc.[4-9]. CrI$_3$ is another 2D van der Waals material. Recently, Huang *et al.* reported that with Curie temperatures ($T_c$) as high as 45 K, ferromagnetism can be retained in monolayer CrI$_3$[10], which is mainly due to the breaking of the Mermin-Wagner theorem[11], where strong magnetic anisotropy counteracts thermal fluctuations, thus stabilizing the long-range magnetic order.

In CrI₃, one Cr ion lies at the center of the octahedron formed by six neighboring iodine ions, and each monolayer is made up of edge-shared $CrI_6$ octahedrons[12]. Upon cooling, CrI₃ undergoes a structural phase transition at approximately 220 K from a high-temperature (HT) monoclinic phase with space group $C2/m$ to a low-temperature (LT) rhombohedral phase $R\bar{3}$ [12-14]. McGuire *et al.* reported that CrI₃ is an insulating ferromagnet with a $T_c$ of 61 K, and every Cr ion has a static magnetic moment of approximately 3.1 $\mu_B$ aligning vertically to each monolayer[12]. The bandgap determined by optical experiments is 1.1-1.2 eV[15-17] and is independent of temperature[16]. Interestingly, the magnetism of CrI₃ is strongly layer dependent: both monolayer and bulk CrI₃ are ferromagnetic (FM)[12] while the bilayer exhibits antiferromagnetic (AFM) interlayer coupling[10,16,18-22]. First-principles calculations demonstrated that the AFM coupling between the bilayers comes from a different stacking order with the $C2/m$ rather than $R\bar{3}$ space group symmetry[23], which is consistent with other research results[24,25]. Furthermore, when different values of the Hubbard $U$ are applied in first-principles calculations, different magnetic states, including AFM and FM states, may be the ground state[23]. Recently, Zhang *et al*. pointed out that there is a close relationship between electronic correlations and magnetism in another van der Waals crystal CrSiTe₃[26]. Therefore, it is crucial to appropriately treat the strong correlation among Cr $3d$ electrons to understand the magnetism of this material.

Although the insulating phase in the FM state has been investigated within the generalized gradient approximation (GGA) and GGA+U schemes for CrI₃[27-30], little research on the insulating phase in the paramagnetic (PM) state has been done. Recently, Craco *et al.* reproduced the PM Mott insulating state through density functional theory in combination with dynamical mean-field theory (DFT+DMFT). Comparing the results obtained by DFT+DMFT calculations with different values of $U$, they concluded that orbital-selective reconstruction plays an important role in this material[31,32]. In addition, McNally *et al.* reported that Hund's coupling $J_H$ determines the charge gap of transition metal materials with half-filled $3d$ orbitals, such as LaMnPO[33]. The octahedral environment of Cr in $CrI_3$ leads to a large crystal field splitting between the Cr $3d$ $t_{2g}$ orbitals and $e_g$ orbitals and results in a half-filled Cr $3d$ $t_{2g}$ orbitals. Despite the similarity of electron filling between $CrI_3$ and LaMnPO, the role of $J_H$ in the charge gap of $CrI_3$ remains unclear. Furthermore, the driving force of the structural phase transition upon cooling has not been explored. To answer these questions, we adopted the DFT+DMFT approach to address the strongly correlated effect of $Cr^{3+}$ $3d$ orbitals and analyzed the electronic structure, atomic histogram and magnetism of $CrI_3$.

## II. Computational details

In this paper, we adopt the experimental lattice constants and atomic positions[12] in all the calculations. To take into account electronic correlation effects, we use the fully charge self-consistent combination of density functional theory and dynamical mean-field theory (DFT+DMFT)[34,35] to theoretically study this material in the PM and FM states. For the DFT part, we use the GGA-PBE exchange correlation functional[36] implemented in the WIEN2K all-electron DFT package[37]. Brillouin zone integrations are performed on a 17×17×17 mesh for the LT phase and an 18×14×18 mesh for the HT phase. The atomic spheres $R_{MT}$ are 2.49 and 2.50 Bohr for Cr and I, respectively, and the plane-wave cutoff $K_{max}$ is given by $R_{MT}K_{max} = 9.0$. To understand the origin of the charge gap in the PM state, both $U = 8$ eV, $J_H = 0$ eV and $U = 5$ eV, $J_H = 0.7$ eV are used in the calculations. The impurity problem in the DFT+DMFT calculation is solved by the continuous time quantum Monte Carlo (CTQMC) method with "exact" double counting, as proposed in reference[38]. In the DMFT impurity solver, we adopt the density-density form of Coulomb repulsion, which speeds up calculation considerably. We break the symmetry between spin up and spin down in the self-energy to simulate the ferromagnetic order of $CrI_3$. To reduce the sign problem in the CTQMC, we use the local axis where the new $x$- and $y$-axes are nearly aligned with the in-plane Cr-I bond direction in an octahedral environment. After achieving the desired accuracy, we perform analytical continuation using the maximum entropy method[35] to obtain the self-energy on the real axis and then calculate the electronic structure. We have checked that spin-orbit coupling affects little on our main findings and shown the band structures obtained with the inclusion of spin-orbit coupling in Fig. S2 in the Supplementary Materials[39]. For simplicity, we report only the results calculated without spin-orbit coupling in the main paper.

## III. Results and discussion

### A. Band structure and density of states

The electronic band structures of CrI$_3$ with the LT crystal structure and HT crystal structure are calculated by the DFT+DMFT method at $T = 50$ K in the PM state (Figs. 1(a)-1(b), (e)-(f)) and FM state (Fig. 1(c)-1(d), 1(g)-1(h)), respectively. Figure 2 presents the corresponding density of states (DOS) of Fig. 1. Since the band structures and DOSs of the HT and LT structures are similar, we focus on the results of the LT structure. The band structure and DOS shown in Fig. 1(a) and Fig. 2(a) are obtained by a large Hubbard $U = 8$ eV and a vanishing Hund's coupling $J_H = 0$ eV. According to experimental results, CrI$_3$ is an insulator even above $T_c$[17,40]. However, the band structure shown in Fig. 1(a) shows metallicity instead of a charge gap in the PM state. The metallicity is characterized by a nonzero DOS around the Fermi level (Fig. 2(a)) which is dominated by contributions from the Cr 3$d$ $t_{2g}$ orbitals and I 5$p$ orbitals. This disagreement between the calculations and experiments illustrates the possibility that CrI$_3$ is not a conventional Mott-Hubbard insulator since a large Hubbard $U = 8$ eV alone cannot open the charge gap.

In previous DMFT calculations, the role of Hund's coupling $J_H$ is highlighted for electronic correlations not only in metallic[41-45] but also insulating systems[33]. To take $J_H$ into account, the other calculations in Fig. 1 and Fig. 2 adopt a more realistic combination of $U = 5$ eV and $J_H = 0.7$ eV, with which the experimental charge gap of CrI$_3$ and the magnetic moment of Cr are very well reproduced. As shown in Fig. 1(b) and Fig. 2(b), when Hund's coupling $J_H = 0.7$ eV and Hubbard $U = 5$ eV are used, CrI$_3$ exhibits insulating characteristics with a charge gap of ~1.1 eV in the PM state, which is in excellent agreement with experimental value[15,46]. It demonstrates that $J_H$ plays an important role in opening the charge gap. Similarly, Chen *et al*. reproduced the charge gap of CrI$_3$ in the paramagnetic state by applying the disordered local moment picture and proved the importance of intra-atomic exchange interaction[47]. We also find that a similar charge gap exists in both majority-spin (Fig. 1(c)) and minority-spin electronic bands (Fig. 1(d)) for CrI$_3$ in the FM state when $U = 5$ eV and $J_H = 0.7$ eV are used, in accordance with previous spin-polarized DFT calculations and experiments[12,27].

In CrI$_3$, the Cr atom has the nominal valence Cr$^{3+}$ ($3d^3$ electronic state) with three electrons occupying the 3$d$ shell, which is different from the calculated Cr 3$d$ orbital occupation number (~4) with approximately three electrons in the $t_{2g}$ orbitals and one electron in the $e_g$ orbitals (Table I). This difference in the orbital occupation number of Cr ions between nominal valence and DFT+DMFT calculations comes from the strong hybridization between Cr 3$d$ $e_g$ orbitals and I 5$p$ orbitals near 3 eV below the Fermi level, as shown in Fig. 2. Furthermore, we find that the static magnetic moment of Cr in the FM state is mainly from the $t_{2g}$ orbitals due to its nearly complete occupation in the majority-spin channel with 2.84 electrons and nearly empty occupation in the minority-spin channel with 0.11 electrons, as shown in the calculated DOSs (Fig. 2(c)-2(d)) and averaged orbital occupation numbers (Table I).

For an isolated atom, the first of Hund's empirical rules states that a maximum total spin $S$ minimizes the total energy. This rule also works for Cr atom in CrI$_3$ solid where the ~3 electrons occupying the Cr 3$d$ $t_{2g}$ orbitals in CrI$_3$ tend to have parallel spins and different angular quantum numbers. It is not favorable for electrons to hop out of or into the Cr 3$d$ $t_{2g}$ orbitals which reduces $S$ and suffers additional energy penalty. As a result, the hopping from and to the Cr 3$d$ $t_{2g}$ orbitals is forbidden and a charge gap opens. We conclude that CrI$_3$ is not a conventional Mott-Hubbard insulator but a Mott-Hund's insulator. As shown in Table SIV in the Supplementary Materials[39], this conclusion is also supported by the fact that the Cr static magnetic moment in the FM state is far more sensitive to $J_H$ than $U$.

Table I. The CTQMC averaged orbital occupation of Cr 3$d$ orbitals from the charge self-consistent DFT+DMFT calculations in the PM and FM states with LT and HT structures at $T = 50$ K. "up" and "dn" denote spin up and spin down electrons,

respectively.

|  | Orbital occupation | | | | | | |
|---|---|---|---|---|---|---|---|
|  | tot | $t_{2g}$ | $e_g$ | $t_{2g}$ up | $t_{2g}$ dn | $e_g$ up | $e_g$ dn |
| LT PM | 4.08 | 2.96 | 1.13 | 1.48 | 1.48 | 0.57 | 0.57 |
| LT FM | 4.09 | 2.95 | 1.13 | 2.84 | 0.11 | 0.76 | 0.37 |
| HT PM | 4.08 | 2.96 | 1.13 | 1.48 | 1.48 | 0.57 | 0.57 |
| HT FM | 4.09 | 2.95 | 1.13 | 2.81 | 0.14 | 0.75 | 0.38 |

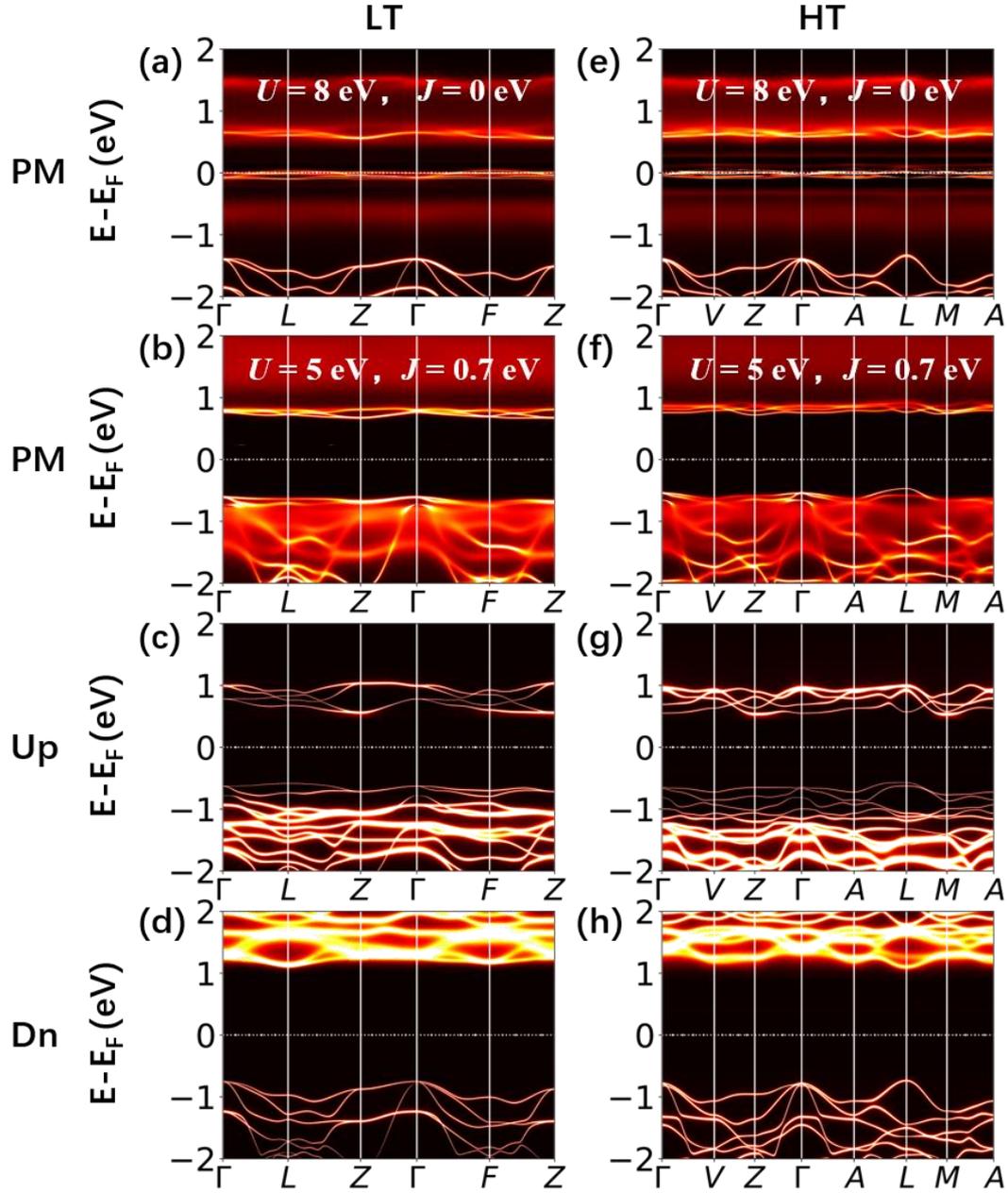

FIG. 1. The DFT+DMFT electronic band structures at $T = 50$ K with a Hubbard $U = 5$ eV and Hund's coupling $J_H = 0.7$ eV except in (a) and (e), where $U = 8$ eV and $J_H = 0$ eV were used. Left (right) column is for LT (HT) structure. The first two rows are for the PM state and the last two rows are for the FM state. "Up" and "Dn" denote spin up and spin down, respectively.

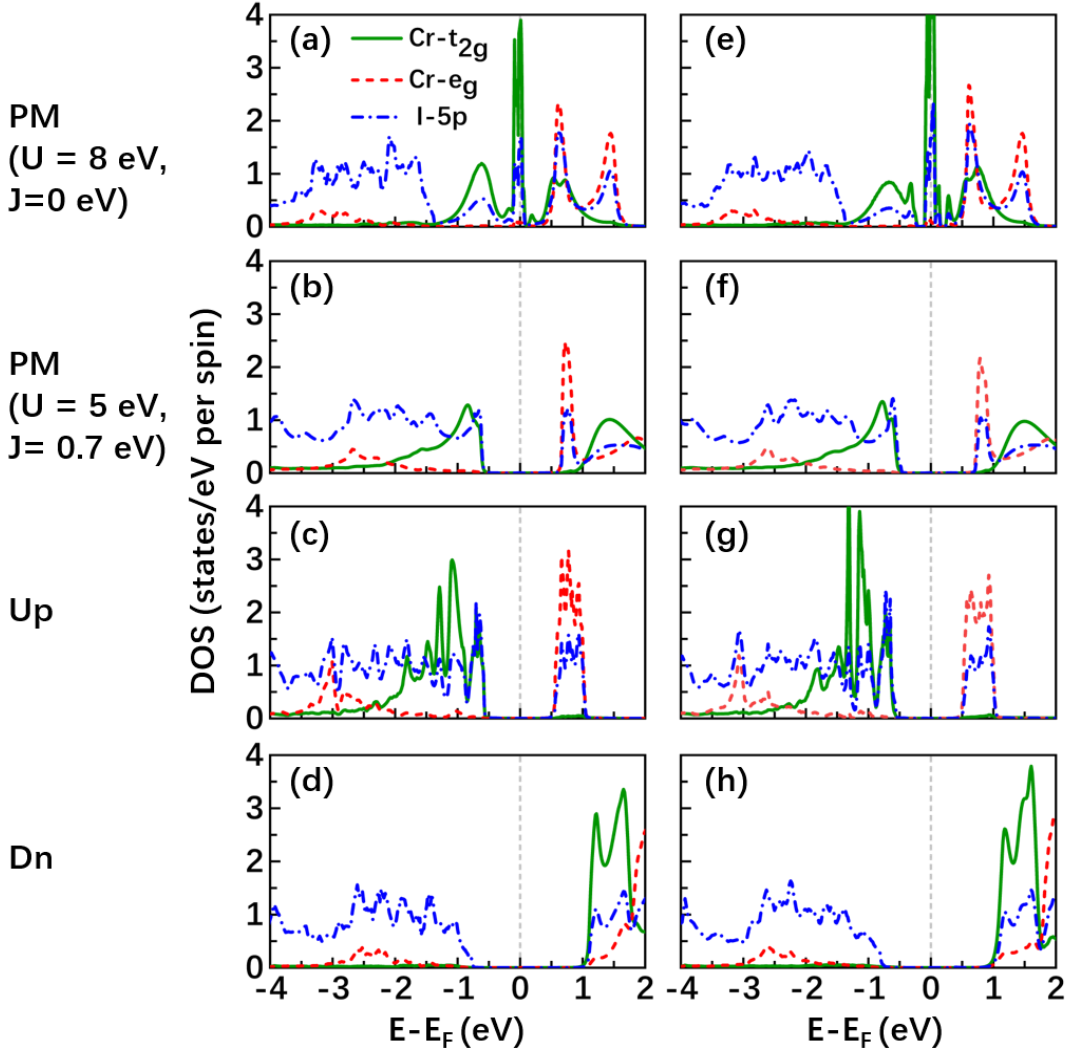

FIG. 2. The DFT+DMFT electronic density of states per unit cell at $T = 50$ K with a Hubbard $U = 5$ eV and Hund's coupling $J_H = 0.7$ eV except in (a) and (e), where $U = 8$ eV and $J_H = 0$ eV were used. Left (right) column is for LT (HT) structure. The first two rows are for the PM state and the last two rows are for the FM state. "Up" and "Dn" denote spin up and spin down, respectively.

## B. Characteristics of Hund's insulators: atomic histogram

Figure 3 presents the Cr $3d$ atomic histogram of the LT structure at 50 K in the PM

state with $U = 8$ eV, $J_H = 0$ eV (Fig. 3(a)) and $U = 5$ eV and $J_H = 0.7$ eV (Fig. 3(b)), and in the FM state with $U = 5$ eV, $J_H = 0.7$ eV (Fig. 3(c)). The atomic histograms of the LT and HT structures at 50 K are almost the same, hence the histogram of the HT structure is not shown. Within the DFT+DMFT scheme, each Cr impurity has 1024 3$d$ states. The atomic histogram refers to the probability of finding a Cr impurity in each atomic state. We note that only those states occupied by $N = 3, 4, 5$ electrons have considerable probabilities, thus the states with larger orbital occupation number ($N > 5$) are not presented in Fig. 3. For the states with the same occupation number $N$, we arrange the states in descending order in accordance with their $|S_z|$ values. As shown in Fig. 3(b), when using Hund's coupling $J_H = 0.7$ eV, only high-spin states have large probabilities. For comparison, when a negligible Hund's coupling $J_H = 0$ eV is used, the probabilities for the high spin states decrease drastically while some low-spin states gain comparable probabilities as shown in Fig. 3(a). It demonstrates that Hund's coupling $J_H$ tends to align the spins of the Cr 3$d$ electrons in parallel.

We now focus on the histograms of the PM and FM states with realistic Hund's coupling $J_H = 0.7$ eV. As shown in In Fig. 3(b) and Fig. 3(c), in both the PM and the FM states, the high-spin states with occupation number $N = 3, 4$, and 5 have the highest probabilities of being occupied, suggesting that there are strong valence fluctuations in CrI$_3$, which share similarities with the histogram of Hund's metals (such as iron pnictides[45]) whereas the valence fluctuations are stronger than those previously reported in the Hund's insulator LaMnPO[33,48]. To trace the origin of the strong valence fluctuations in CrI$_3$, we check all the 3$d$ states (1024 states) and sort them according to the configuration of the $t_{2g}$ orbitals (explained in detail in the Supplementary Materials[39]). As shown in Table SIII in the Supplementary Materials[39], the electrons in the $t_{2g}$ orbitals tend to have parallel spins, whereas the electrons of $e_g$ orbitals are arranged randomly. Therefore, Cr 3$d$ $e_g$ orbitals are responsible for the strong valence fluctuations in CrI$_3$.

The strong fluctuations in the $e_g$ orbitals give rise to a negligible spin moment of the $e_g$ electrons in the FM state. The occupation number of Cr 3$d$ orbitals calculated within the DFT+DMFT scheme is approximately 4 (Table I), out of which three are for $t_{2g}$ orbitals and one is for $e_g$ orbitals. Therefore, the DFT+DMFT calculated static magnetic moment of ~3.1 $\mu_B$ comes mainly from the $t_{2g}$ orbitals due to Hund's rule and is in excellent agreement with the experimental value of ~3.1 $\mu_B$ at low temperature[12].

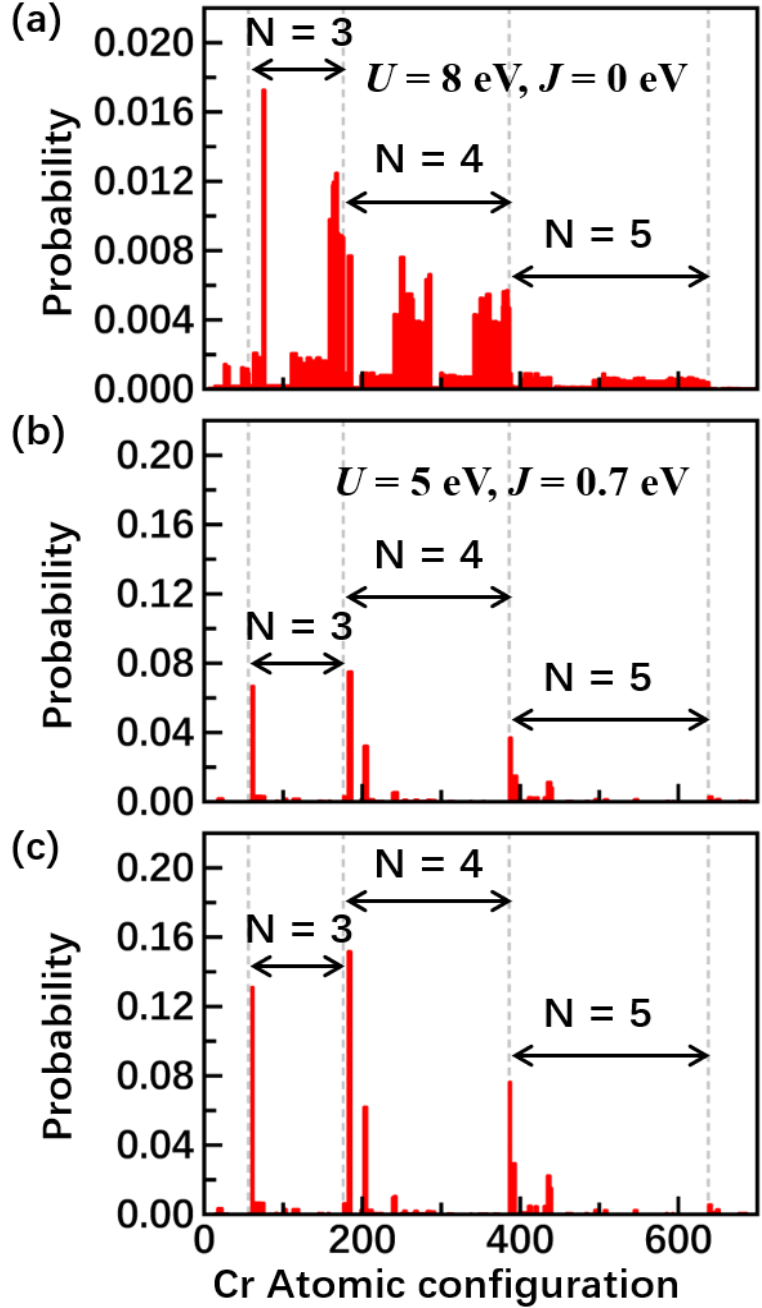

FIG. 3. DFT+DMFT calculated atomic histogram of Cr $3d$ orbitals in CrI$_3$ with the LT structure: (a) PM state with $U = 8$ eV and $J_H = 0$ eV. (b) PM state and (c) FM state with $U = 5$ eV and $J_H = 0.7$ eV. The probability scale of (a) is only one tenth of (b) and (c). The states with occupation number exceeding five are not shown due to their negligible

probabilities.

## C. Magnetism and structural phase transition at finite temperature

Figure 4 presents the Cr magnetic moment versus $T$ obtained by the DFT+DMFT calculations for both the LT and HT structures. The fluctuating magnetic moment is nearly independent of temperature or structure (~3.1 $\mu_B$ per Cr for the LT structure and HT structure at both 50 K and 160 K), so only the fluctuating magnetic moments of the LT structure at various temperatures are shown for clarity. The fluctuating magnetic moment is calculated according to the formula $<m_z> = 2\sum_i P_i |S_z|_i$, where $i$ is the index of the 1024 atomic states, $P_i$ and $|S_z|_i$ are the corresponding probability and absolute value of the total spin, respectively. Consequently, the value of the fluctuating magnetic moment is the upper limit of the static magnetic moment in the magnetically ordered states. The fluctuating magnetic moment is equal to the static magnetic moment of the ferromagnetic order only if quantum fluctuations vanish. As shown in Fig. 4, the fluctuating magnetic moment hardly changes with decreasing temperature. It is expected that the fluctuating magnetic moment is ~3.1 $\mu_B$ at very low temperatures. At 50 K, the calculated static magnetic moment in the ferromagnetic state is already very close to the fluctuating magnetic moment, suggesting that the ferromagnetic static magnetic moment is ~3.1 $\mu_B$ at 2 K, which agrees well with the experimental value[12].

Our DFT calculated static magnetic moments of the LT and HT structures in the FM state are the same, ~3.0 $\mu_B$ per Cr, which is consistent with previous calculated results[12]. In contrast, as shown in Fig. 5, the DFT+DMFT calculated FM static magnetic moments of the LT and HT structures exhibit clear differences at finite temperatures, where the static magnetic moment of the HT structure drops much faster than that of the LT structure with increasing temperature. As a result, the calculated Curie temperature $T_c$ (the temperature corresponding to the disappearance of the static magnetic moment) of the HT structure is approximately 116 K, while the static magnetic moment of the LT structure remains at ~2.4 $\mu_B$ per Cr at the same temperature. Therefore, we conclude that the structural differences between the HT and LT structures originating from stacking pattern have little effect on the fluctuating magnetic moment but is essential to the static magnetic moment in the FM state, especially at high temperatures. We note that the $T_c$ calculated from the DFT+DMFT method is generally higher than the experimentally determined value ($T_c \sim 61$ K), which is not surprising since DFT+DMFT calculations neglect spatial fluctuations of the magnetic order.

Based on the Heisenberg model $H = -J \sum_{l,\delta} \hat{s}_l \cdot \hat{s}_{l+\delta}$ we can estimate the magnetic coupling energy of systems with different magnetic orders. In CrI$_3$, the structural differences between the LT and HT phases mainly originate from different stacking orders, which indicates that there are some differences in the interlayer magnetic coupling parameters $J_2$, whereas the intralayer coupling parameters $J_1$ are almost identical for the two phases. Previous calculations[28] and experiments[49] demonstrate that the magnetic coupling parameter $J_1$ is much larger than $J_2$ in CrI$_3$ and two other chromium halides including CrCl$_3$[50] and CrBr$_3$[51-53] by means of spin wave analysis. Therefore, the interlayer magnetic coupling is negligible, which shows that the magnetic coupling energy of the Heisenberg model depends chiefly on the intralayer coupling $J_1$. In Fig. 4, we find that the FM Cr static magnetic moment of the LT phase is always larger than that of the HT phase at the same temperature. According to the Heisenberg model, the LT structure is more stable with lower energy because it has a larger static magnetic moment and almost the same nearest-neighbor exchange interaction parameter as the HT phase. For example, the magnetic coupling energy of the LT phase is ~2.6 times lower than that of the HT phase at 100 K. Regarding the structural phase transition from the HT phase to the LT phase with decreasing temperature in CrI$_3$, the magnetic coupling energy greatly contributes to the difference in the total energy between the two phases. Therefore, we propose that the structural phase transition from the HT to the LT phase is driven by ferromagnetic spin fluctuations.

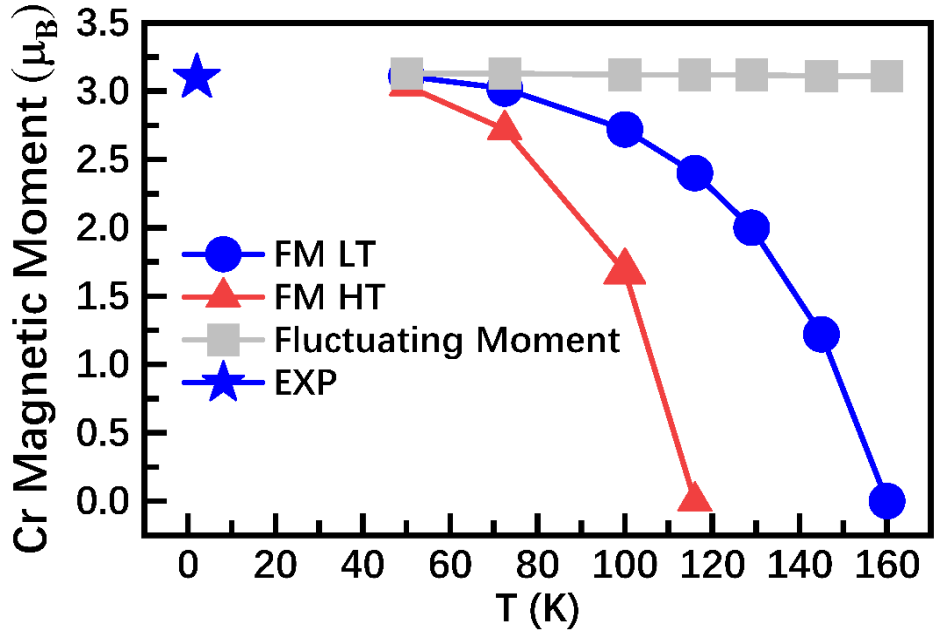

FIG. 4. The DFT+DMFT calculated magnetic moments of Cr at different temperatures for both the LT and HT structures. The blue circle and red triangle represent the values of the ferromagnetic static magnetic moments of the LT and HT structures, respectively. The calculated fluctuating magnetic moment of the LT structure is shown as gray square. The experimentally determined static magnetic moment[12] at $T$ = 2 K is shown as blue star.

### D. Two other Hund's insulators: $CrCl_3$ and $CrBr_3$

In addition to $CrI_3$, we have carried out DFT+DMFT calculations of the electronic structures of two other chromium halides $CrCl_3$ and $CrBr_3$ in the $R\bar{3}$ phases using both $U$ = 8 eV, $J_H$ = 0 eV and $U$ = 5 eV, $J_H$ = 0.7 eV. The lattice constants and atomic positions of $CrCl_3$ are taken from Ref. [54], while the lattice constants of $CrBr_3$ are taken from Ref. [55]. The atomic positions of $CrBr_3$ are not provided in Ref. 55 hence are fully optimized using the Vienna *ab-initio* simulation package (VASP)[56] until all the force components become smaller than 0.04 eV/Å, which is consistent with previous calculations[31,32]. We come to the same conclusion that Hund's coupling is decisive for the charge gap in the PM state in both $CrCl_3$ and $CrBr_3$. As shown in Figs. 5(a) and

5(c), despite a large $U$ value of 8 eV, there are several bands crossing the Fermi level when Hund's coupling is absent, i.e., $J_H = 0$. In contrast, there are charge gaps in the band structures obtained by using $U = 5$ eV and $J_H = 0.7$ eV, as shown in Figs. 5(b) and 5(d). Consequently, our conclusion that CrI$_3$ is a Mott-Hund's insulator rather than a Mott-Hubbard insulator also applies to CrCl$_3$ and CrBr$_3$. In Fig. 6, we present a comparison of the DFT+DMFT calculated DOSs of CrCl$_3$, CrBr$_3$ and CrI$_3$ obtained using $U = 5$ eV and $J_H = 0.7$ eV in the PM state. From CrCl$_3$ to CrI$_3$, we find that the band gap gets smaller, and the peaks of DOSs of the Cr $t_{2g}$ and $e_g$ orbitals above the Fermi level get broader. Below the Fermi level, the energy region of the Cr $t_{2g}$ orbitals hybridizing with the halogen element $p$ orbitals gets closer to the Fermi level as shown in Figs. 6(a)-(c).

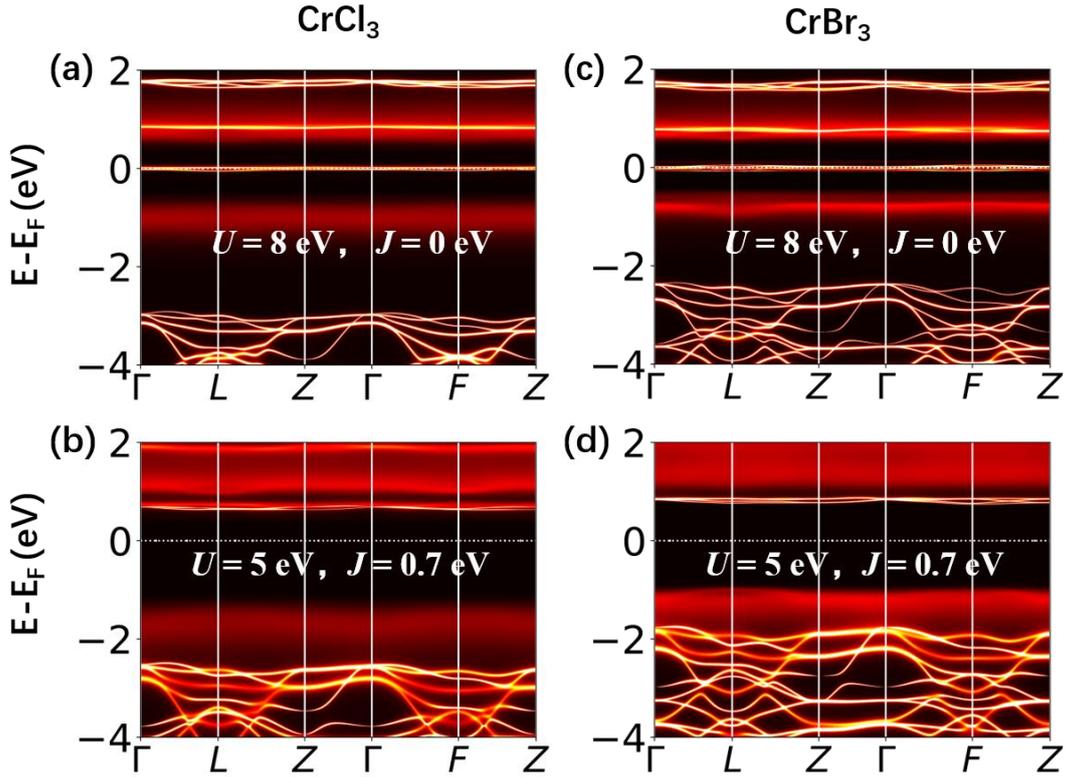

FIG. 5. The DFT+DMFT calculated electronic band structures at $T = 50$ K with $U=8$ eV, $J_H= 0$ eV (a, c) and $U=5$ eV, $J_H=0.7$ eV (b, d) of CrCl$_3$ (a, b) and CrBr$_3$ (c, d) in the paramagnetic state. Both CrCl$_3$ and CrBr$_3$ are in the $R\bar{3}$ phase.

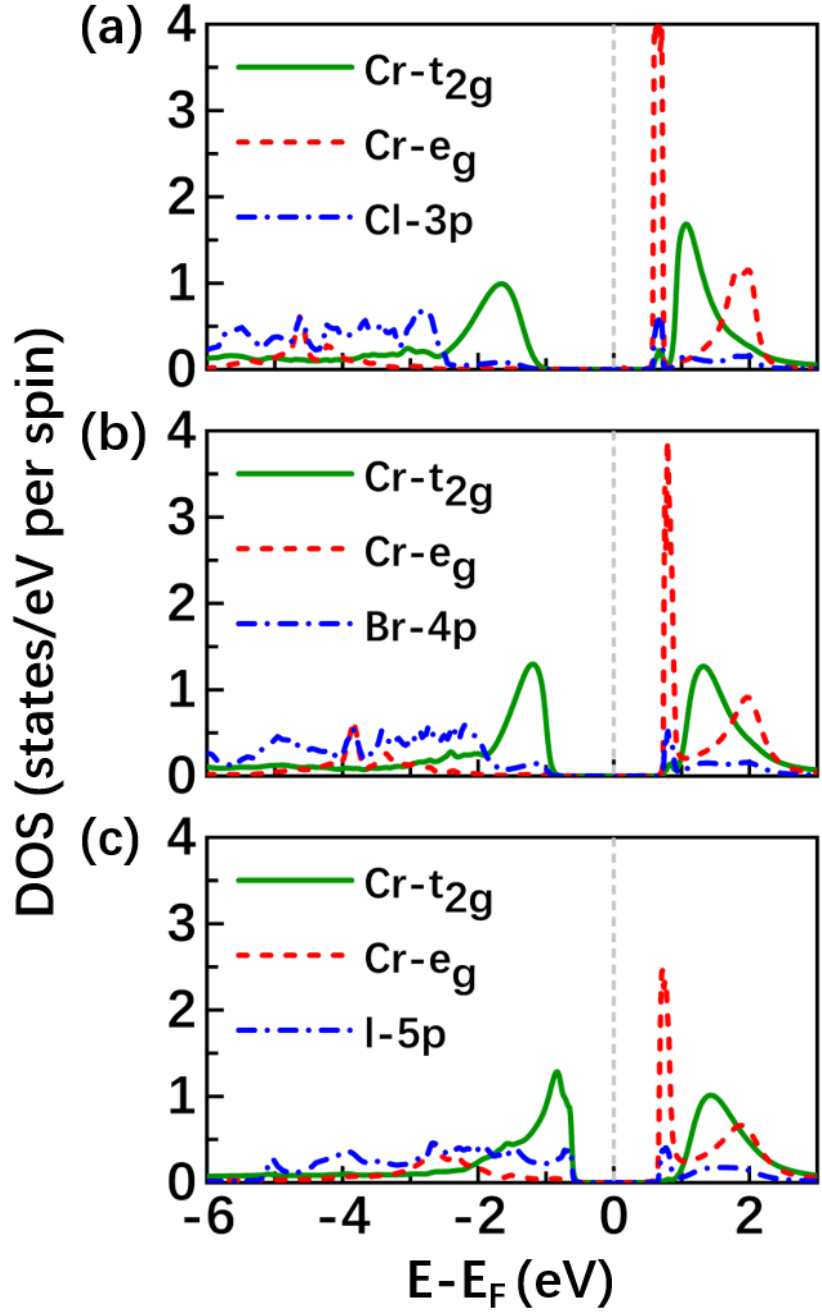

FIG. 6. The DFT+DMFT calculated electronic density of states per unit cell in the PM state at $T = 50$ K with $U = 5$ eV, $J_H = 0.7$ eV of (a) $CrCl_3$, (b) $CrBr_3$ and (c) $CrI_3$ in the $R\bar{3}$ phase.

## IV. Conclusions

Our DFT+DMFT calculations indicate that $CrI_3$ is a Mott-Hund's insulator rather than a conventional Mott-Hubbard insulator. Including a realistic Hund's coupling $J_H$~0.7 eV is essential to open a charge gap of ~1.1 eV in the PM state. This conclusion also applies to other chromium halides including CrCl$_3$ and CrBr$_3$. By analyzing the atomic histogram of the Cr 3$d$ shell, we find that the strong valence fluctuations in CrI$_3$ come mainly from Cr $e_g$ orbitals, unlike the valence fluctuations in the previously observed Mott-Hund's insulator LaMnPO where all the 3$d$ orbitals are responsible for the valence fluctuations[33]. The calculated static magnetic moment of Cr ions in the FM state in the LT $R\bar{3}$ phase is larger than that in the HT $C2/m$ phase at the same temperature. Therefore, the $R\bar{3}$ phase is more stable with a lower magnetic coupling energy at low temperature, which suggests that the driving force of the structural phase transition from the $C2/m$ phase to the $R\bar{3}$ phase upon cooling is ferromagnetic spin fluctuations.

**Acknowledgments**

This work was supported by the National Natural Science Foundation of China (Grant No. 12074041 and 11674030), the Foundation of the National Key Laboratory of Shock Wave and Detonation Physics (Grant No. 6142A03191005), the National Key Research and Development Program of China through Contract No. 2016YFA0302300, and the start-up funding of Beijing Normal University. The calculations were carried out with high performance computing cluster of Beijing Normal University in Zhuhai.

*Requests for materials should be addressed to Z.P.Y. at yinzhiping@bnu.edu.cn.

# Supplementary Materials for "Electronic structure and magnetism of the Hund's insulator CrI$_3$"

Tianye Yu, Rui Liu, Huican Mao, Xiaobo Ma, Guangwei Wang, Zhihong Yuan, Pengyu Zheng, Yiran Peng, and Zhiping Yin[*]

Department of Physics and Center for Advanced Quantum Studies, Beijing Normal University, Beijing 100875, China

Table SI. Charge gap (in unit of eV) of CrI$_3$ in the paramagnetic state calculated by DFT+DMFT using different $U$ and $J_H$ at temperature of 50 K.

| $J_H$/$U$ (eV) | 4 | 5 | 6 |
|---|---|---|---|
| 0.6 | 1.04 | 1.09 | 1.09 |
| 0.7 | 1.10 | 1.12 | 1.14 |
| 0.8 | 1.11 | 1.13 | 1.14 |

Table SII. Static magnetic moment (in unit of $\mu_B$) of Cr in the ferromagnetic state calculated by DFT+DMFT using different $U$ and $J_H$ at temperature of 50 K.

| $J_H$/$U$ (eV) | 4 | 5 | 6 |
|---|---|---|---|
| 0.6 | 2.75 | 2.90 | 2.97 |
| 0.7 | 3.09 | 3.11 | 3.13 |
| 0.8 | 3.18 | 3.19 | 3.20 |

Tables SI and SII present the dependences of the paramagnetic charge gap and ferromagnetic static magnetic moment of Cr on different combinations of $U$ and $J_H$ in the $R\bar{3}$ phase [1] at the temperature of 50 K, obtained by the DFT+DMFT method[2,3]. We find that $U$=5.0 eV, $J_H$=0.7 eV can reproduce the experimental charge gap (1.1-1.2 eV[4-6]) and static magnetic moment of Cr (3.1 $\mu_B$[1]) very well thus the combination of $U$=5.0 eV and $J_H$=0.7 eV were used in the DFT+DMFT calculations reported in the main text.

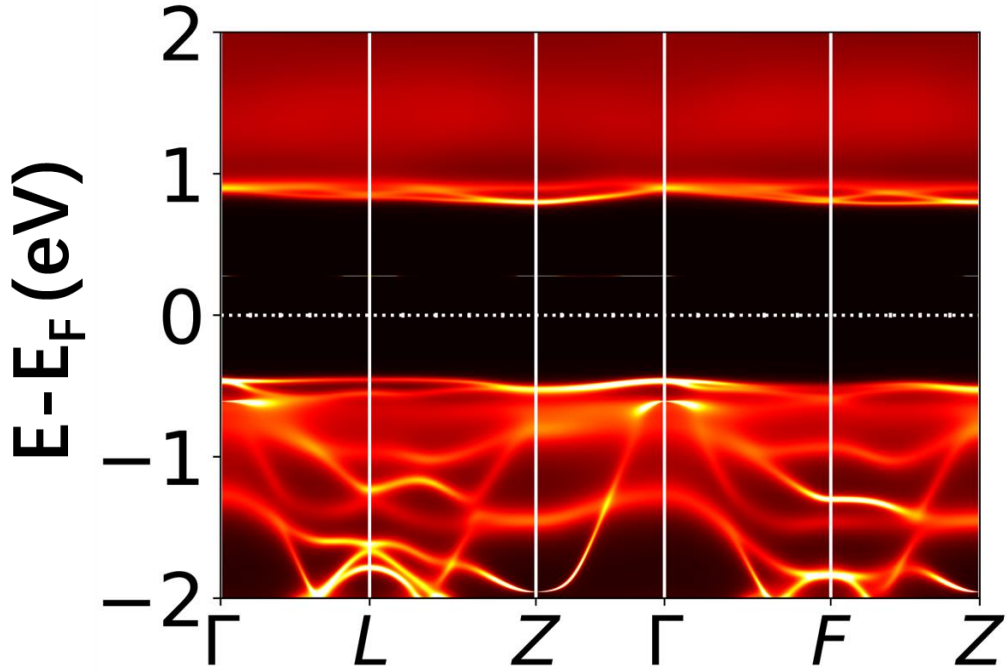

FIG. S1. The DFT+DMFT calculated electronic band structure in the paramagnetic state at $T=50$ K with Hubbard $U=2$ eV and Hund's coupling $J_H=0.7$ eV.

We carry out DFT+DMFT calculations with a small $U=2$ eV and the same $J_H=0.7$ eV in order to clarify the importance of $J_H$ for opening the charge gap. As shown in Fig. S1, the DFT+DMFT calculated band structure presents a charge gap of ~1.0 eV. It confirms that it is the Hund's rule $J_H$ rather than the Hubbard $U$ that is responsible for opening the charge gap in $CrI_3$.

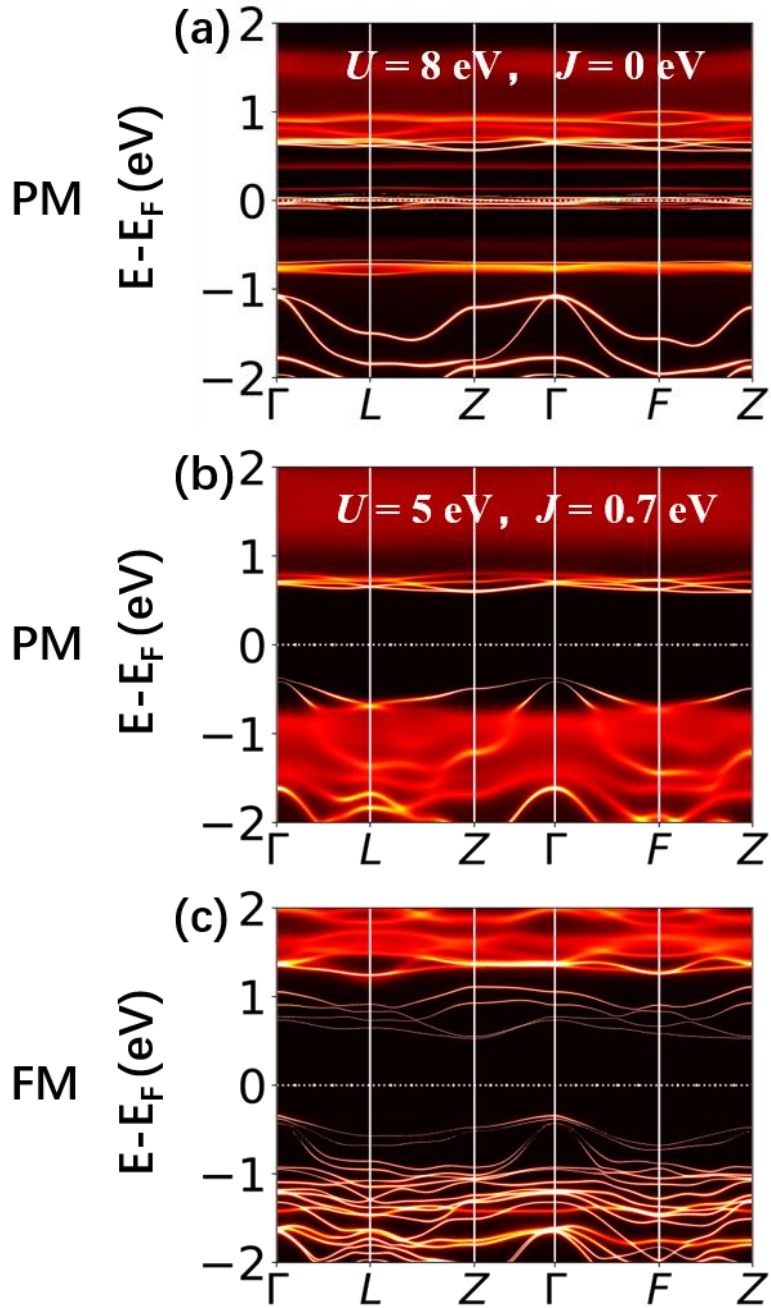

FIG. S2. The DFT+DMFT calculated electronic band structures at $T=50$ K with a Hubbard $U=5$ eV and Hund's coupling $J_H=0.7$ eV except in (a), where $U=8$ eV and $J_H=0$ eV are used. Figs. S2(a)-(b) are obtained in the paramagnetic state whereas (c) is obtained in the ferromagnetic state. Spin-orbit coupling is taken into account in all the three calculations.

Figure S2 presents the DFT+DMFT calculated electronic band structures when spin-orbit coupling is taken into account. We find that there are several bands crossing the

Fermi level when $J_H$ is absent as shown in Fig. S2(a), however there is a charge gap when $J_H$=0.7 eV is used as shown in Fig. S2(b). Therefore, we conclude that spin-orbit coupling does not affect our conclusion that CrI$_3$ is a Mott-Hund's insulator. Moreover, the inclusion of spin-orbit coupling hardly affects the calculated fluctuating magnetic moment in the paramagnetic state or the calculated static magnetic moment of Cr in the ferromagnetic state. The band structure obtained in the ferromagnetic state with the inclusion of spin-orbit coupling is shown in Fig. S2(c).

Table SIII. The Cr atomic configurations with probabilities greater than 0.01 and their corresponding probabilities calculated by DFT+DMFT in the PM state at 50 K. The first three orbitals represent the $t_{2g}$ orbitals, and the last two represent the $e_g$ orbitals. "u" and "d" denote single occupation with one electron in the spin up or spin down orientation, respectively, and 0 and 2 denote empty and double occupation, respectively.

| Probability | Atomic configuration |
|---|---|
| 0.0666 | uuu00 |
| 0.0666 | ddd00 |
| 0.0749 | uuuu0 |
| 0.0749 | dddd0 |
| 0.0749 | uuu0u |
| 0.0749 | ddd0d |
| 0.0320 | uuud0 |
| 0.0320 | dddu0 |
| 0.0320 | uuu0d |
| 0.0320 | ddd0u |
| 0.0368 | uuuuu |
| 0.0368 | ddddd |
| 0.0147 | uuuud |
| 0.0147 | uuudu |
| 0.0147 | ddddu |
| 0.0147 | dddud |
| 0.0111 | uuu20 |
| 0.0111 | ddd20 |
| 0.0111 | uuu02 |
| 0.0111 | ddd02 |

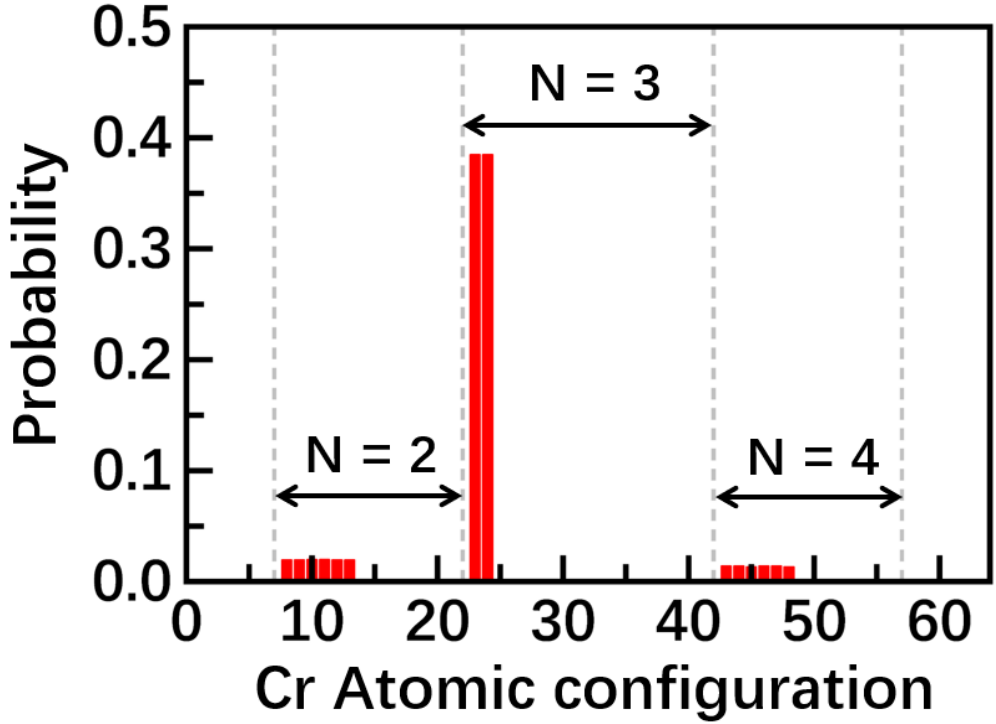

FIG. S3. The Cr atomic histogram of the Cr-$3d$ shell in CrI$_3$ obtained at 50 K in the PM state with the LT structure. Note that all the 1024 atomic configurations in the calculation are divided into 64 groups according to the configurations of $t_{2g}$ orbitals. The probability refers to the sum of the probabilities of 16 states within each group, and the number $N$ presented in the figure is the orbital occupation number of $t_{2g}$ orbitals.

To understand the origin of the strong valence fluctuations in CrI$_3$, we list all the Cr $3d$ states of Fig. 3b in the main text with a probability greater than 0.01 in Table SIII. Each of the five $3d$ orbitals of the Cr atom has four possible occupied configurations, including u, d, 0, and 2. Here, u and d denote the half-occupied configuration with one spin-up and spin-down electron, respectively. 0 and 2 denote empty and double occupation, respectively. As a result, there are $4^5 = 1024$ configurations for the Cr $3d$ shell. In Table SIII, the first three orbitals are $t_{2g}$, and the last two orbitals are $e_g$. We notice that the favorable states with large probabilities have a common feature: the $t_{2g}$ orbitals are always half filled by electrons with parallel spins, whereas the spin arrangement of electrons occupying the $e_g$ orbitals is random. Therefore, the strong valence fluctuations are mainly contributed by the $e_g$ orbitals instead of all five $3d$ orbitals. To demonstrate this, we divide all the 1024 states into 64 groups according to the electronic configuration of $t_{2g}$ orbitals and sum up the probabilities of 16 states (all possible configurations of $e_g$ orbitals) within each group. The sum of probabilities for each group is provided in Fig. S3. There are no strong valence fluctuations because only two states, the uuu and ddd states, where the three electrons occupying $t_{2g}$ orbitals are all spin-up or spin-down, hold considerable probabilities. Therefore, it is concluded that the strong valence fluctuations shown in Fig. 3(b) in the main text are mainly derived from $e_g$ orbitals.

Table SIV. Static magnetic moment of Cr in the FM state of CrI$_3$ calculated by DFT+DMFT using different $U$ and $J_H$ values. The fitted values are calculated by the simple formula $M = (0.04\ U + 1.41\ J_H + 1.90\ eV)\ \mu_B / eV$.

| $U$ (eV) | $J_H$ (eV) | Static magnetic moment ($\mu_B$) | Fitted value ($\mu_B$) |
|---|---|---|---|
| 10 | 0.6 | 3.10 | 3.11 |
| 9 | 0.6 | 3.08 | 3.08 |
| 8 | 0.6 | 3.06 | 3.04 |
| 5 | 0.8 | 3.19 | 3.21 |
| 5 | 0.7 | 3.11 | 3.07 |
| 5 | 0.6 | 2.90 | 2.93 |

We obtain a simple formula to describe the FM Cr static magnetic moment $M = (0.04\ U + 1.41\ J_H + 1.90\ eV)\ \mu_B / eV$ in terms of $U$ and $J_H$ by fitting the calculated static magnetic moments with the corresponding $U$ and $J_H$, as listed in Table SIV. This formula indicates that $J_H$ is much more important in determining the static magnetic moment than $U$. This is a key characteristic of a Hund's insulator.